\documentclass[10pt,final,letterpaper]{IEEEtran}
\usepackage{amsmath}
\usepackage{latexsym}
\usepackage{graphicx}
\usepackage{bbding}
\usepackage{indentfirst}
\usepackage{cases}
\usepackage{algorithm2e}
\usepackage{subeqnarray}
\IEEEoverridecommandlockouts

\begin{document}
\title{Performance Analysis for Physical Layer Security in Multi-Antenna Downlink Networks with Limited CSI Feedback}
\author{\authorblockN{Xiaoming~Chen and Rui~Yin
\thanks{This work was supported by the grants from the NUAA Research
Funding (No. NN2012004), the open research fund of National Mobile
Communications Research Laboratory, Southeast University (No.
2012D16) and the Doctoral Fund of Ministry of Education of China
(No. 20123218120022).}
\thanks{Xiaoming~Chen (e-mail: {\tt chenxiaoming@nuaa.edu.cn})
is with the College of Electronic and Information Engineering,
Nanjing University of Aeronautics and Astronautics, and also with
the National Mobile Communications Research Laboratory, Southeast
University, China. Rui~Yin is with the Department of Information
Science and Electronic Engineering, Zhejiang University, China.}}}
\maketitle

\markboth{IEEE Wireless Communications Letters, Vol. XX, No. Y,
Month 2013}{} \maketitle

\begin{abstract}
Channel state information (CSI) at the transmitter is of importance
to the performance of physical layer security based on multi-antenna
networks. Specifically, CSI is not only beneficial to improve the
capacity of the legitimate channel, but also can be used to degrade
the performance of the eavesdropper channel. Thus, the secrecy rate
increases accordingly. This letter focuses on the quantitative
analysis of the ergodic secrecy sum-rate in terms of feedback amount
of the CSI from the legitimate users in multiuser multi-antenna
downlink networks. Furthermore, the asymptotic characteristics of
the ergodic secrecy sum-rate in two extreme cases is investigated in
some detail. Finally, our theoretical claims are confirmed by the
numerical results.
\end{abstract}


\section{Introduction}
Due to the open nature of wireless channel, information transmission
security is always a critical issue in wireless communications.
Traditionally, secure communication is realized by using
cryptography technology. With the development of interception
technology, cryptography technology becomes more and more complex,
resulting in high computation. In fact, it has been proven by
information theory that wireless security can be guaranteed by
physical layer technology, namely physical layer security
\cite{Wyner} \cite{PLS1}. The advantage of physical layer security
lies in that it is independent of the interception ability of the
eavesdropper, so it is appealing in secure communications with
low-complexity nodes.

The performance of physical layer security depends on the difference
between the legitimate channel capacity and the eavesdropper channel
capacity, namely secrecy rate \cite{Multiantenna1}. Intuitively,
multi-antenna techniques can improve the legitimate channel capacity
and degrade the performance of the eavesdropper channel
simultaneously by exploiting the spatial degree of freedom. Thus,
physical layer security based on multi-antenna system draws
considerably attentions \cite{Multiantenna2} \cite{Multiantenna3}.
In \cite{Multiantenna4}, the problem regarding the maximum secrecy
rate in multiple-antenna system was addressed by designing an
optimal transmit beam, assuming that full CSI related to the
legitimate and eavesdropper channels is available at the
transmitter. However, in practical systems, the eavesdropper is
usually passive and hidden, so the transmitter is difficult to
obtain the eavesdropper CSI. Under such a condition, it seems
impossible to provide a steady secrecy rate, and thus the ergodic
secrecy rate is adopted as a useful and intuitive metric to evaluate
security. In \cite{ErgodicRate1}, the ergodic secrecy rate for
Gaussian MISO wiretap channels was analyzed and the corresponding
optimal transmit beam was presented. The work shown in
\cite{ErgodicRate2} considered the scenario where all nodes have
multiple antennas, and the ergodic rate for such an MIMO secrecy
scenario has been developed. Another advantage of the multi-antenna
secrecy system lies in that it supports concurrent transmission of
multiple legitimate users. On one hand, multiuser transmission can
improve the ergodic secrecy sum-rate. On the other hand, the
inter-user interference can weaken the interception effect. In
\cite{ErgodicRate3}, the ergodic secrecy sum-rate for orthogonal
random beamforming with opportunistic scheduling was presented. Yet,
in multiuser downlink networks, orthogonal random beamforming may
suffer performance loss if the number of users is not so large. In
\cite{CSIquantization}, the authors proposed to convey the quantized
CSI from the legitimate user to the transmitter for transmit beam
design in a single receiver secure scenario.

Combining the advantages of multiuser transmission and CSI
conveyance, this letter proposes to perform limited feedback
zero-forcing (ZFBF) at the transmitter in multiuser multi-antenna
downlink networks \cite{ZFBF}. We focus on the quantitative analysis
of the ergodic secrecy sum-rate, and derive its closed-form
expression in terms of the feedback amount. Furthermore, the
asymptotic characteristics of the ergodic secrecy sum-rate is
analyzed, so as to provide an explicit insight on system parameter
optimization for physical layer security in multiuser multi-antenna
downlink networks.

The rest of this letter is organized as follows. We first provide an
overview of the multiuser secure system model in Section II, and
then the ergodic secrecy sum-rate is derived in Section III. We
investigate the asymptotic characteristics of the ergodic secrecy
sum-rate in Section IV. Some numerical results are given to show the
accuracy of the theoretical analysis in Section V. Finally, the
whole letter is concluded in Section VI.

\section{System Model}
\begin{figure}[h] \centering
\includegraphics [width=0.4\textwidth] {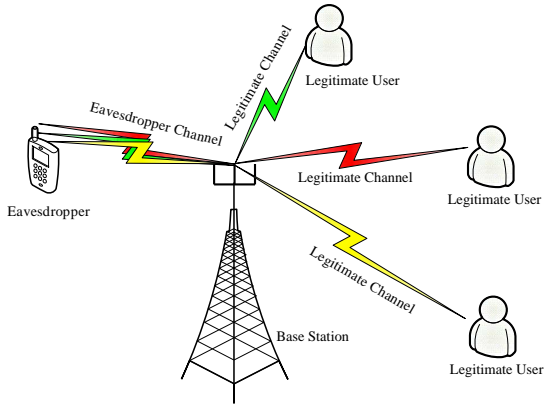}
\caption {An overview of the multi-antenna downlink network
employing physical layer security.} \label{Fig1}
\end{figure}

We consider a homogeneous multiuser multi-antenna downlink network
employing physical layer security for secure communication, where a
base station (BS) equipped with $N_t>1$ antennas communicates with
$K$ single antenna legitimate users (LU), while a passive single
antenna eavesdropper attempts to intercept the transmission
information, as shown in Fig.\ref{Fig1}. In this letter, we only
consider the case of $K=N_t$, since the multi-antenna system can
admit $N_t$ users at most for each time slot. Note that if $K>N_t$,
user scheduling can be adopted to select $N_t$ users. We use
$\textbf{h}_k$ to denote the $N_t$ dimensional legitimate channel
vector from the BS to the $k$th LU with independent and identically
distributed (i.i.d.) zero mean and unit variance complex Gaussian
entries. In addition, we use $\alpha\textbf{g}$ to denote the $N_t$
dimensional eavesdropper channel vector from the BS to the
eavesdropper, where $\alpha$ is the relative path loss defined as
the ratio of path loss of the eavesdropper channel and that of the
legitimate channel, and $\textbf{g}$ is the channel fast fading
distributed as $\mathcal{CN}(0,\textbf{I}_{N_t})$.

The whole network is operated in slotted time. At the beginning of
each time slot, the LU first estimates the CSI related to its
legitimate channel, and then chooses an optimal codeword to quantize
the CSI from a predetermined quantization codebook
$\mathcal{H}_{k}=\{\hat{\textbf{h}}_{k,1},\cdots,\hat{\textbf{h}}_{k,2^B}\}$
of size $2^B$ according to the following selection criterion by
assuming perfect CSI at the LU:
\begin{equation}
\hat{\textbf{h}}_{k,opt}=\arg\max\limits_{\hat{\textbf{h}}_{k,i}\in\mathcal{H}_{k}}
\left|\tilde{\textbf{h}}_k^H\hat{\textbf{h}}_{k,i}\right|^2\label{eqn1}
\end{equation}
where $\tilde{\textbf{h}}_k=\frac{\textbf{h}_k}{\|\textbf{h}_k\|}$
is the channel direction vector. Next, the index of the optimal
codeword is conveyed by the $k$th LU and $\hat{\textbf{h}}_{k,opt}$
is recovered at the BS from the same codebook as the instantaneous
CSI of the $k$th legitimate channel. Based on the $N_t$ LUs'
feedback information, the BS determines the optimal transmit beam
$\textbf{w}_k$ for the $k$th LU by making use of zero-forcing
beamforming (ZFBF) design method. Specifically, given
$\hat{\textbf{h}}_{k,opt}$, we first construct its complementary
matrix
\begin{eqnarray}
\hat{\textbf{H}}_k=[\hat{\textbf{h}}_{1,opt},\cdots,\hat{\textbf{h}}_{k-1,opt},
\hat{\textbf{h}}_{k+1,opt},\cdots,\hat{\textbf{h}}_{N_t,opt}]\nonumber
\end{eqnarray}
Taking singular value decomposition (SVD) to $\hat{\textbf{H}}_k$,
if $\textbf{V}_{k}^{\perp}$ is the matrix composed of the right
singular vectors with zero singular values, then we randomly choose
a unit norm vector from the space spanned by
$\textbf{V}_{k}^{\perp}$ as the transmit beam $\textbf{w}_k$. Since
$\textbf{V}_{k}^{\perp}$ is the null space of
$\hat{\textbf{H}}_{k}$, we have
\begin{equation}
\hat{\textbf{h}}_{i,opt}^H\textbf{w}_k=0, i\neq k\label{eqn2}
\end{equation}
Thus, the receive signals at the $k$th LU and the eavesdropper can
be expressed as
\begin{equation}
y_k=\sqrt{P}\textbf{h}_k^H\textbf{w}_kx_k+\sqrt{P}\textbf{h}_k^H\sum\limits_{i=1,i\neq
k}^{N_t}\textbf{w}_ix_i+n_k\label{eqn3}
\end{equation}
and
\begin{equation}
y_0=\sqrt{P}\alpha\textbf{g}\sum\limits_{i=1}^{N_t}\textbf{w}_ix_i+n_0\label{eqn4}
\end{equation}
respectively, where $x_k$ is the normalized Gaussian distributed
transmit signal for the $k$th LU, $P$ is the transmit power, $n_k$
is the additive Gaussian white noise with variance $\sigma^2$. In
this context, the ergodic secrecy sum-rate can be expressed as
\begin{eqnarray}
R&=&\sum\limits_{k=1}^{N_t}E\left[\log_2\left(1+\gamma_k\right)-\log_2\left(1+\zeta_k\right)\right]\nonumber\\
&=&\sum\limits_{k=1}^{N_t}E\left[\log_2\left(1+\gamma_k\right)\right]-\sum\limits_{k=1}^{N_t}E\left[\log_2\left(1+\zeta_k\right)\right]\label{eqn5}
\end{eqnarray}
where
$\gamma_k=\frac{|\textbf{h}_k^H\textbf{w}_k|^2}{\sum\limits_{i=1,i\neq
k}^{N_t}|\textbf{h}_k^H\textbf{w}_i|^2+\sigma^2/P}$ and
$\zeta_k=\frac{|\textbf{g}^H\textbf{w}_k|^2}{\sum\limits_{i=1,i\neq
k}^{N_t}|\textbf{g}^H\textbf{w}_i|^2+\sigma^2/\alpha^2P}$ are the
SINR related to the $k$th LU's signal at the $i$th LU and the
eavesdropper. (\ref{eqn5}) follows the fact that the legitimate and
eavesdropper channels are independent of each other.

The focus of this letter is on the quantitative analysis of the
ergodic secrecy sum-rate in terms of feedback amount $B$ and BS
antenna number (LU number) $N_t$.

\section{Performance Analysis of Physical Layer Security}
In this section, we intend to derive the ergodic secrecy sum-rate
$R$ in a multiuser multi-antenna downlink network in presence of a
passive eavesdropper. As seen in (\ref{eqn5}), the key of computing
the ergodic secrecy sum-rate is to obtain the distributions of the
SINRs $\gamma_k$ and $\zeta_k$. In what follows, we first give an
investigation of their distributions employing limited feedback
ZFBF. According to the theory of random vector quantization (RVQ)
\cite{RVQ}, the relationship between the original and the quantized
channel direction vectors is given by
\begin{equation}
\tilde{\textbf{h}}_k=\sqrt{1-a}\hat{\textbf{h}}_{k,opt}+\sqrt{a}\textbf{s}_k\label{eqn6}
\end{equation}
where
$a=\sin^2\left(\angle(\tilde{\textbf{h}}_k,\hat{\textbf{h}}_{k,opt})\right)$
is the magnitude of the quantization error, and $\textbf{s}_k$ is an
unit norm vector isotropically distributed in the nullspace of
$\hat{\textbf{h}}_{k,opt}$, and is independent of $a$.

Substituting (\ref{eqn6}) into the definition of the SINR
$\gamma_k$, we have
\begin{eqnarray}
\gamma_k&=&\frac{|\textbf{h}_k^H\textbf{w}_k|^2}{a\|\textbf{h}_k\|\sum\limits_{i=1,i\neq
k}^{N_t}|\textbf{s}_k^H\textbf{w}_i|^2+\sigma^2/P}\label{eqn7}\\
&\stackrel{d}{=}&\frac{\chi^2_2}{\Gamma(N_t-1,\delta)\sum\limits_{i=1,i\neq
k}^{N_t}\beta(1,N_t-2)+\sigma^2/P}\label{eqn8}
\end{eqnarray}
\begin{eqnarray}
&\stackrel{d}{=}&\frac{\chi^2_2}{\delta\chi^2_{2(N_t-1)}+\sigma^2/P}\label{eqn9}
\end{eqnarray}
where $\stackrel{d}{=}$ denotes the equality in distribution, and
$\delta=2^{-\frac{B}{N_t-1}}$. Eq. (\ref{eqn7}) holds true by
applying the property of limited feedback ZFBF in (\ref{eqn2}). In
(\ref{eqn8}), $\textbf{w}_k$ is designed regardless of
$\textbf{h}_k$, so $|\textbf{h}_k^H\textbf{w}_k|^2$ is $\chi^2_2$
distributed with the probability density function (pdf) of
$\exp(-x)$. According to the theory of quantization cell
approximation \cite{QCA}, the product of channel gain
$\|\textbf{h}_k\|^2$ and the magnitude of quantization error $a$ is
$\Gamma(N_t-1,\delta)$ distributed with the pdf of
$\frac{1}{\delta^{N_t-1}(N_t-2)!}x^{N_t-2}\exp(-\frac{x}{\delta})$.
Additionally, since $\textbf{s}_k$ and $\textbf{w}_i$ are i.i.d.
isotropic vectors in the $N_t-1$ dimensional null space of
$\hat{\textbf{h}}_{k,opt}$, $|\textbf{s}_k^H\textbf{w}_i|^2$ is
$\beta(1,N_t-2)$ distributed with pdf of $(N_t-2)(1-x)^{N_t-3}$
\cite{RVQ}. In (\ref{eqn9}), the product of a $\Gamma(N_t-1,\delta)$
distributed random variable and a $\beta(1,N_t-2)$ distributed
random variable is $\delta\chi^2_2$, so the sum of $N_t-1$
independent $\delta\chi^2_2$ distributed random variables is
$\delta\chi^2_{2(N_t-1)}$ distributed, where the pdf of the
$\delta\chi^2_{2(N_t-1)}$ distribution is
$\frac{x^{N_t-1}\exp(-x)}{\Gamma(N_t-1)}$. Let
$y\sim\chi^2_{2(N_t-1)}$ and $z\sim\chi^2_2$, the cumulative
distribution function (cdf) of $\gamma_k$ is derived as
\begin{eqnarray}
F(x)\!\!\!\!\!&=&P_r\left(\frac{z}{\delta y+\sigma^2/P}\leq x\right)\nonumber\\
&=&\int_0^{\infty}F_{Z|Y}(x(\delta y+\sigma^2/P))f_Y(y)dy\nonumber\\
&=&\int_0^{\infty}\left(1-\exp\left(-x(\delta
y+\frac{\sigma^2}{P})\right)\right)\frac{y^{N_t-1}\exp(-y)}{\Gamma(N_t-1)}dy\nonumber\\
&=&1-\frac{\exp(-x\sigma^2/P)}{(1+\delta
x)^{N_t-1}}\nonumber\\
&*&\int_0^{\infty}\frac{\exp(-(1+\delta x)y)((1+\delta
x)y)^{N_t-1}}{\Gamma(N_t-1)}d((1+\delta x)y)\nonumber\\
&=&1-\frac{\exp(-x\sigma^2/P)}{(1+\delta x)^{N_t-1}}\label{eqn10}
\end{eqnarray}
where $F_{Z|Y}(\cdot)$ is the conditional cdf of $z$ for a given
$y$, $f_Y(\cdot)$ is the probability density function (pdf) of $y$,
and $\Gamma(\cdot)$ is the Gamma function. (\ref{eqn10}) holds true
since $\frac{\exp(-(1+\delta x)y)((1+\delta
x)y)^{N_t-1}}{\Gamma(N_t-1)}$ is the pdf of $(1+\delta x)y$.

Examining the definition of $\zeta_k$, it can be considered as a
special $\gamma_k$ without CSI feedback, namely $B=0$ or $\delta=1$,
so the cdf of $\zeta_k$ can be derived based on (\ref{eqn10}) as
\begin{equation}
G(x)=1-\frac{\exp(-x\sigma^2/\alpha^2P)}{(1+x)^{N_t-1}}\label{eqn11}
\end{equation}
Substituting (\ref{eqn10}) and (\ref{eqn11}) into (\ref{eqn5}), we
have
\begin{eqnarray}
R\!\!\!\!\!&=&N_t\int_0^{\infty}\left(\log_2(1+x)\left(F^{'}(x)-G^{'}(x)\right)\right)dx\nonumber\\
&=&N_t\log_2(e)\int_0^{\infty}\left(\ln(1+x)\left(F^{'}(x)-G^{'}(x)\right)\right)dx\nonumber\\
&=&N_t\log_2(e)\int_0^{\infty}\left(\frac{1-F(x)}{1+x}-\frac{1-G(x)}{1+x}\right)dx\nonumber\\
&=&\frac{N_t\log_2(e)}{\delta^{N_t-1}}\int_0^{\infty}\frac{\exp(-x\sigma^2/P)}{(x+1)(x+\delta^{-1})^{N_t-1}}dx\nonumber\\
&-&N_t\log_2(e)\int_0^{\infty}\frac{\exp(-x\sigma^2/\alpha^2P)}{(x+1)^{N_t}}dx\nonumber\\
&=&\frac{N_t\log_2(e)}{\delta^{N_t-1}}I_1\left(\sigma^2/P,\delta^{-1},N_t-1\right)\nonumber
\end{eqnarray}
\begin{eqnarray}
&-&\frac{N_t\log_2(e)}{\Gamma(N_t)}\bigg(\sum\limits_{m=1}^{N_t-1}\Gamma(m)\left(-\frac{\sigma^2}{P\alpha^2}\right)^{N_t-1-m}\nonumber\\
&-&\left(-\frac{\sigma^2}{P\alpha^2}\right)^{N_t-1}\exp\left(\frac{\sigma^2}{P\alpha^2}\right)\textmd{Ei}\left(-\frac{\sigma^2}{P\alpha^2}\right)\bigg)\label{eqn12}
\end{eqnarray}
where
\begin{eqnarray}
I_1(x,y,z)&=&\int_{0}^{\infty}\frac{\exp\left(-xt\right)}{(t+1)(t+y)^z}dt\nonumber\\
&=&\sum\limits_{i=1}^{z}(-1)^{i-1}(1-y)^{-i}I_2(x,y,
z-i+1)\nonumber\\
&+&(y-1)^{-z}I_2(x,1,1)\nonumber
\end{eqnarray}
\begin{equation}
I_2(x,y,z)=\left\{\begin{array}{ll} \exp(xy)\textmd{Ei}(xy) & z=1\\
\sum\limits_{k=1}^{z-1}\frac{(k-1)!}{(z-1)!}\frac{(-x)^{z-k-1}}{y^k}\\
+\frac{(-x)^{z-1}}{(z-1)!}\exp(xy)\textmd{Ei}(xy) & z\geq2
\end{array}\right.\nonumber
\end{equation}
and $\textmd{Ei}(z)=\int_{-\infty}^{z}\frac{\exp(t)}{t}dt$ is the
exponential-integral function. $F^{'}(z)$ denotes the derivative of
$F(z)$ with respect to $z$. The first term of (\ref{eqn12}) is
derived following \cite{Modeselection} and the second term is
obtained according to [Eq. 3.3532, 15] Thus, we obtain the ergodic
secrecy sum-rate as a function of feedback amount $B$, antenna
number $N_t$, channel condition $\alpha$ and transmit power $P$, so
as to provide an insight on parameter selection in secure
communications.

\section{Asymptotic Performance Analysis}
In this section, for the sake of evaluating the performance easily,
we analyze the asymptotic characteristics of ergodic secrecy
sum-rate in some extreme cases.

\subsection{Interference-Limited Case}
If transmit power $P$ is quite high or LU number is so large, the
receive noise is negligible with respect to the inter-user
interference. Thus, the SINRs are reduced as
$\gamma_k=\frac{|\textbf{h}_k^H\textbf{w}_k|^2}{\sum\limits_{i=1,i\neq
k}^{N_t}|\textbf{h}_k^H\textbf{w}_i|^2}$ and
$\zeta_k=\frac{|\textbf{g}^H\textbf{w}_k|^2}{\sum\limits_{i=1,i\neq
k}^{N_t}|\textbf{g}^H\textbf{w}_i|^2}$, respectively. In this case,
the cdfs can be expressed as
\begin{equation}
F(x)=1-\frac{1}{(1+\delta x)^{N_t-1}}\label{eqn13}
\end{equation}
and
\begin{equation}
G(x)=1-\frac{1}{(1+x)^{N_t-1}}\label{eqn14}
\end{equation}
Similar to (\ref{eqn12}), the ergodic secrecy sum-rate can be
computed as
\begin{eqnarray}
R&=&\frac{N_t\log_2(e)}{\delta^{N_t-1}}\int_0^{\infty}\frac{1}{(x+1)(x+\delta^{-1})^{N_t-1}}dx\nonumber\\
&-&N_t\log_2(e)\int_0^{\infty}\frac{1}{(x+1)^{N_t}}dx\nonumber\\
&=&N_t\log_2(e)B(1,N_t-1)_2F_1\left(N_t-1,1;N_t;1-\delta\right)\nonumber\\
&-&N_t\log_2(e)/(N_t-1)\label{eqn15}
\end{eqnarray}
where $B(x,y)$ is the Beta function and
$_2F_1(\alpha,\beta;\gamma;z)$ is the Gauss hypergeometric function.
Eq. (\ref{eqn15}) is derived according to [eq. 3.1971, 15]. It is
found that in interference-limited case, the ergodic secrecy
sum-rate is independent of transmit power $P$ and channel condition
$\alpha$. Given $N_t$ and $B$, the ergodic secrecy sum-rate is a
constant.

\subsection{Noise-Limited Case}
If the inter-user interference is negligible due to low transmit
power or strong noise, the SINRs can be approximated as
$\gamma_k=\frac{P}{\sigma^2}|\textbf{h}_k^H\textbf{w}_k|^2$ and
$\zeta_k=\frac{P\alpha^2}{\sigma^2}|\textbf{g}^H\textbf{w}_k|^2$,
respectively. As analyzed earlier, both
$|\textbf{h}_k^H\textbf{w}_k|^2$ and $|\textbf{g}^H\textbf{w}_k|^2$
are $\chi_2^2$ distributed, so the cdfs of the SINRs are given by
\begin{equation}
F(x)=1-\exp\left(-\sigma^2/Px\right)\label{eqn16}
\end{equation}
and
\begin{equation}
G(x)=1-\exp\left(-\sigma^2/(P\alpha^2)x\right)\label{eqn17}
\end{equation}
In this case, the ergodic secrecy sum-rate can be computed as
\begin{eqnarray}
R&=&N_t\log_2(e)\big(\exp(\sigma^2/P)\textmd{Ei}(\sigma^2/P)\nonumber\\
&-&\exp(\sigma^2/(P\alpha^2))\textmd{E}{\textmd{i}}(\sigma^2/(P\alpha^2))\big)\label{eqn18}
\end{eqnarray}
It is found that the ergodic secrecy sum-rate in noise-limited case
is independent of feedback amount. In other words, the ergodic
secrecy sum-rates with different $B$s asymptotically approach the
same value as $P$ decreases.

\section{Numerical Results}
To examine the accuracy of the derived ergodic secrecy sum-rate in
multiuser multi-antenna downlink networks, we present several
numerical results in the following scenario: we set $N_t=5$ and
$B=4$. In addition, we define SNR$=10\log_{10}\frac{P}{\sigma^2}$ as
the transmit SNR.

\begin{figure}[h] \centering
\includegraphics [width=0.5\textwidth] {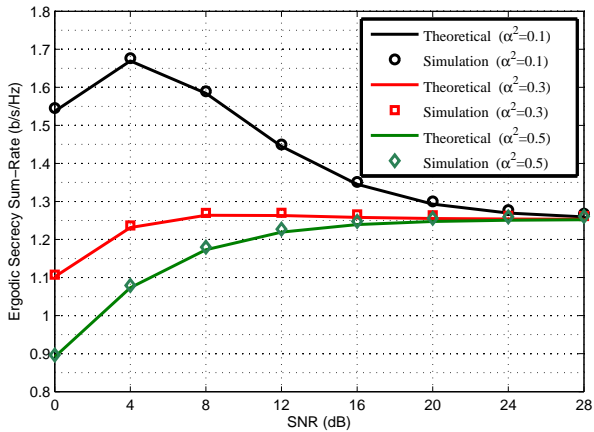}
\caption {Comparison of theoretical and simulation results.}
\label{Fig2}
\end{figure}

Fig.\ref{Fig2} compares the ergodic secrecy sum-rates obtained by
theoretical analysis and numerical simulation with different
relative path losses $\alpha$. It is seen that the theoretical
result is quite consistent with the simulation result in the whole
SNR region, which validate the high accuracy of the derived result.
Note that relative path loss $\alpha$ has a great impact on the
ergodic secrecy sum-rate in the low SNR region, since the small
$\alpha$, indicating the weak intercepting signal at the
eavesdropper, leads to a large ergodic secrecy sum-rate. With the
increase of SNR, the ergodic secrecy sum-rates with different
$\alpha$s asymptotically approach the same value, which is
equivalent to the interference-limited case whose ergodic secrecy
sum-rate is independent of relative path loss and SNR, as shown in
(\ref{eqn15}). Moreover, it is found that the ergodic secrecy
sum-rate is not an increasing function of SNR, since both the
legitimate and eavesdropper channel capacities increase as SNR adds.
Thus, there is an optimal SNR in the sense of maximizing the ergodic
secrecy sum-rate, which provides an important insight on system
parameter selection, especially in green communications that
concentrate on energy efficiency.

\section{Conclusion}
This letter focused on the performance analysis for physical layer
security in multiuser multi-antenna downlink networks with quantized
CSI feedback, and derived the closed-form expression of ergodic
secrecy sum-rate in terms of feedback amount and channel condition.
Through asymptotic analysis, we obtained the relatively simple
expression of ergodic secrecy sum-rate in two extreme scenarios.
Numerical simulation reconfirmed the high accuracy of the derived
theoretical results.


\begin{thebibliography}{1}

\bibitem{Wyner}
A. D. Wyner, ``The wire-tap channel," \emph{Bell Syst. Tech. J.},
vol. 54, pp. 1355-1387, Oct. 1975.

\bibitem{PLS1}
P. K. Gopala, L. Lai, and H. El. Gamal, ``On the secrecy capacity of
fading channels," \emph{IEEE Trans. Inf. Theory}, vol. 54, no. 10,
pp. 4687-4698, Oct. 2008.

\bibitem{Multiantenna1}
T. Liu, and S. Shamai, ``A note on the secrecy capacity of the
multiple-antenna wiretap channel," \emph{IEEE Trans. Inf. Theory},
vol. 55, no. 6, pp. 2547-2553, Jun. 2009.

\bibitem{Multiantenna2}
A. Khisti, and G. W. Wornell, ``Secure transmission with multiple
antenna-part II: the MIMOME wiretap channel,"  \emph{IEEE Trans.
Inf. Theory}, vol. 56, no. 11, pp. 5515-5532, Nov. 2010.

\bibitem{Multiantenna3}
X. Chen, and L. Lei, ``Energy-efficient optimization for physical
layer security in multi-antenna downlink networks with QoS
guarantee," \emph{IEEE Commun. Lett.}, vol. 17, no. 4, pp. 637-640,
Apr. 2013.

\bibitem{Multiantenna4}
A. Khisti, and G. W. Wornell, ``Secure transmission with multiple
antennas I: the MIMOME wiretap channel," \emph{IEEE Trans. Inf.
Theory}, vol. 56, no. 7, pp. 3088-3104, Jul. 2010.

\bibitem{ErgodicRate1}
J. Li, and A. P. Petropulu, ``On ergodic secrecy rate for Gaussian
MISO wiretap channels," \emph{IEEE Trans. Wireless Commun.}, vol.
10, no. 4, pp. 1176-1187, Apr. 2011.

\bibitem{ErgodicRate2}
Z. Ding, M. Peng, and H-H. Chen, ``A general relaying transmission
protocol for MIMO secrecy communications," \emph{IEEE Trans.
Commun.}, vol. 60, no. 11, pp. 3461-3471, Nov. 2012.

\bibitem{ErgodicRate3}
I. Krikidis, and B. Ottersten, ``Secrecy sum-rate for orthogonal
random beamforming with opportunistic scheduling," \emph{IEEE Signal
Process. Lett.}, vol. 20, no. 2, pp. 141-144, Feb. 2013.

\bibitem{CSIquantization}
Bashar, Shafi, Zhi Ding, and Geoffrey Ye Li, ``On secrecy of
codebook-based transmission beamforming under receiver limited
feedback," \emph{IEEE Trans. Wireless Commun.},  vol. 10, no. 4, pp.
1212-1223, Apr. 2011.

\bibitem{ZFBF}
X. Chen, and C. Yuen, ``Efficient resource allocation in rateless
coded MU-MIMO cognitive radio network with QoS provisioning and
limited feedback," \emph{IEEE Trans. Veh. Tech.}, vol. 62, no. 1,
pp. 395-399, Jan. 2013.

\bibitem{RVQ}
N. Jindal, ``MIMO Broadcast channels with finite-rate feedback,"
\emph{IEEE Trans. Info. Theory}, vol. 52, no. 11, pp. 5045-5060,
Nov. 2006.

\bibitem{QCA}
K. K. Mukkavilli, A. Sabharwal, E. Erkip, and B. Aazhang, ``On
beamforming with finite rate feedback in multiple-antenna systems,"
\emph{IEEE Trans. Inf. Theory}, vol. 49, no. 10, pp. 2562-2579, Oct.
2003.

\bibitem{Modeselection}
J. Zhang, M. Kountouris, J. G. Andrews, and R. W. Heath Jr.,
``Multi-mode transmission for the MIMO broadcast channel with
imperfect channel state information," \emph{IEEE Trans. Commun.},
vol. 59, no. 3, pp. 803-814, Mar. 2011.

\bibitem{Mathbook}
I. S. Gradshteyn, and I. M. Ryzhik, ``Tables of intergrals, series,
and products," \emph{Acedemic Press}, USA, 2007.

\end{thebibliography}
\end{document}